\documentstyle[manuscript,aps]{revtex}
\begin{document}
\draft
\title{The averaged tensors of the relative energy-momentum and angular
momentum in general relativity and some of their applications}
\author{Janusz Garecki}
\address{Institute of Physics, University of Szczecin, Wielkopolska 15; 70-451
Szczecin, POLAND\footnote{e-mail: garecki@sus.univ.szczecin.pl}}
\date{\today}
\maketitle
\begin{abstract}
There exist different kinds of averaging of the
differences of the energy-momentum and angular momentum in normal
coordinates {\bf NC(P)} which give tensorial quantities. The obtained
averaged quantities are equivalent mathematically because they differ
only by constant scalar dimensional factors.
One of these averaging was used in our papers [1-8] giving the {\it canonical
superenergy and angular supermomentum tensors}.

In this paper we present another averaging of the differences of the energy-momentum and
angular momentum which gives tensorial quantities with proper dimensions of the
energy-momentum and angular momentum densities. But these averaged
relative energy-momentum and angular momentum tensors, closely
related to the canonical superenergy and angular supermomentum
tensors, {\it depend on some fundamental length $L>0$}.

The averaged relative energy-momentum and angular momentum tensors of the
gravitational field obtained in the paper can be applied,
like the canonical superenergy and angular supermomentum tensors,
to {\it coordinate independent} analysis (local and in special cases also
global)  of this field.

We have applied the averaged relative energy-momentum tensors to
analyze vacuum gravitational energy and momentum and to analyze energy
and momentum of the Friedman (and also more general) universes. The obtained results
are interesting, e.g., the averaged relative energy density is {\it positive
definite} for the all Friedman universes.
\end {abstract}

\section{The averaged relative energy-momentum and angular momentum
tensors in general relativity}
In the papers [1-8] we have defined the canonical superenergy and
angular supermomentum tensors, matter and gravitation, in general
relativity ({\bf GR}) and studied their properties and physical
applications. In the case of the gravitational field these tensors gave
us some substitutes of the non-existing gravitational
energy-momentum and gravitational angular momentum tensors.

The canonical superenergy and angular supermomentum tensors were
obtained pointwise as a result of some special averaging of the
differences of the energy-momentum and angular momentum in normal
coordinates {\bf NC(P)}. The role of the normal coordinates {\bf NC(P)}
is, of course, auxilliary, only to extract tensorial quantities even
from pseudotensorial ones.

The dimensions of the components of the canonical superenergy and
angular supermomentum tensors can be written
down as:[the dimensions of the components of an energy-momentum or
angular momentum tensor (or pseudotensor)]$\times m^{-2}$.

In this paper we propose a new averaging of the energy-momentum and
angular momentum differences in {\bf NC(P)} which is very like to the
averaging used in [1-8] and which gives the averaged
quantities with proper dimensionality of the energy-momentum and angular
momentum densities.

Namely, we propose the following general definition of the averaged tensor
(or pseudotensor)  $T_a^b$
\begin{equation}
<{\it T_a^{~b}}(P)> := \displaystyle\lim_{\varepsilon\to
0}{\int\limits_{\Omega}{\bigl[{\it T_{(a)}^{~~~(b)}}(y) - {\it
T_{(a)}^{~~~(b)}}(P)\bigr]
d\Omega}\over\varepsilon^2/2\int\limits_{\Omega}d\Omega},
\end{equation}
where
\begin{equation}
{\it T_{(a)}^{~~~(b)}}(y) := T_i^{~k}(y){}e^i_{~(a)}(y){}e_k^{~(b)}(y),
\end{equation}
\begin{equation}
{\it T_{(a)}^{~~~(b)}}(P):= T_i^{~k}(P){}e^i_{~(a)}(P){}e_k^{~(b)}(P) = {\it T_a^{~b}}(P)
\end{equation}
are the tetrad (or physical) components of a tensor or a pseudotensor
$T_i^{~k}(y)$ which describes an energy-momentum distribution, $y$ is the collection of normal coordinates {\bf NC(P)}
at a given point {\bf P}, $e^i_{~(a)}(y),~e_k^{~(b)}(y)$ denote an
orthonormal tetrad field and its dual, respectively,
\begin{equation}
e^i_{~(a)}(P) = \delta^i_a,~e_k^{~(a)}(P)
=\delta^a_k,~e^i_{~(a)}(y)e_i^{~(b)}(y) =\delta_a^b,
\end{equation}
and they are parallelly propagated along geodesics through {\bf P}.

For a sufficiently small domain $\Omega$ which surrounds {\bf P} we
require
\begin{equation}
\int\limits_{\Omega}{y^id\Omega} = 0,~~\int\limits_{\Omega}{y^iy^kd\Omega} =
\delta^{ik} M,
\end{equation}
where
\begin{equation}
M = \int\limits_{\Omega}{(y^0)^2 d\Omega} = \int\limits_{\Omega} {(y^1)^2
d\Omega} = \int\limits_{\Omega}{(y^2)^2 d\Omega} =
\int\limits_{\Omega}{(y^3)^2 d\Omega},
\end{equation}
is a common value of the moments of inertia of the domain $\Omega$ with
respect to the subspaces $y^i = 0, ~~(i = 0,1,2,3)$.

The procedure of averaging of an energy-momentum tensor or an
energy-momentum pseudotensor given in (1) is a four-dimensional
modification of the proposition by Mashhoon [9-12].

Let us choose $\Omega$ as a small analytic ball defined by
\begin{equation}
(y^0)^2 + (y^1)^2 + (y^2)^2 + (y^3)^2 \leq R^2 = \varepsilon^2 L^2,
\end{equation}
which can be described in a covariant way in terms of the auxiliary
positive-definite metric $h^{ik} := 2v^iv^k - g^{ik}$, where $v^i$ are
the components of the four-velocity of an observer {\bf O} at rest at
{\bf P} (see, e.g., [1-8]). $\varepsilon$ means a small parameter: $\varepsilon\in (0;1)$
and $L >0$ is a fundamental length.

Since at {\bf P} the tetrad and normal components are equal, from now on
we will write the components of any quantity at {\bf P} without (tetrad)
brackets, e.g., $T_a^{~b}(P)$ instead of $T_{(a)}^{~~~(b)}(P)$ and so on.

Let us now make the following expansions for the energy-momentum tensor
of matter $T_i^{~k}(y)$ and for $e^i_{~(a)}(y), e_k^{~(b)}(y)$ [13]
\begin{eqnarray}
T_i^{~k}(y) &=& {\hat T}_i^{~k} + \nabla_l {\hat T}_i^{~k}y^l +1/2{{\hat
T}_i^{~k}}{}_{,lm} y^ly^m + R_3\nonumber\\
&=& {\hat T}_i^{~k} +\nabla_l{\hat T}_i^{~k} y^l
+1/2\biggl[\nabla_{(l}\nabla_{m)}{\hat T}_i^{~k}\nonumber\\
&-&1/3{\hat R}^c_{~(l\vert i\vert m)}{}{\hat T}_c^{~k} +1/3 {\hat
R}^k_{~(l\vert c\vert m)}{} {\hat T}_i^{~c}\biggr] y^ly^m + R_3,
\end{eqnarray}
\begin{equation}
e^i_{~(a)}(y) = {\hat e}^i_{~(a)} + 1/6{\hat R}^i_{~lkm}{\hat
e}^k_{~(a)} y^ly^m + R_3,
\end{equation}
\begin{equation}
e_k^{~(b)}(y) = {\hat e}_k^{~(b)} -1/6{\hat R}^p_{~lkm}{\hat e}_p^{~(b)}
y^ly^m + R_3,
\end{equation}
which give (1) in the form
\begin{equation}
<_m T_a^{~b}(P)> =\displaystyle\lim_{\varepsilon\to
0}{\int\limits_{\Omega}\bigl[\nabla_l{\hat T}_a^{~b} y^l +
1/2\nabla_{(l}\nabla_{m)}{\hat T}_a^{~b}y^ly^m + THO\bigr]d\Omega\over
\varepsilon^2/2\int\limits_{\Omega}d\Omega},
\end{equation}
where $THO$ means the terms of higher order in the expansion of the
differences $T_{(a)}^{~~~(b)}(y) - T_{(a)}^{~~~(b)}(P) =
T_{(a)}^{~~~(b)}(y) - T_a^{~b}(P)$; $R_3$ is the remainder of the third
order and  $\nabla$ denotes covariant differentiation. Hat denotes
the value of an object at {\bf P} and the round brackets denote
symmetrization from which the indices inside vertical lines, e.g.,
$(a\vert c\vert b)$ are excluded.

The first and $THO$ terms in the numerator of (11) do not contribute to
$<_m T_a^{~b}(P)>$. Hence, we finally get from (11)
\begin{equation}
<_m T_a^{~b}(P)> = _m S_a^{~b}(P) {L^2\over 6},
\end{equation}
where
\begin{equation}
_m S_a^{~b}(P) := \delta^{lm}\nabla_{(l}\nabla_{m)}{\hat T}_a^{~b}
\end{equation}
is the {\it canonical superenergy tensor of matter} [1-8].

By introducing the four velocity ${\hat v}^l \dot = \delta^l_0,~v^lv_l =1$ of
an observer {\bf O} at rest at {\bf P} and the local metric ${\hat
g}^{ab}\dot = \eta^{ab}$, where $\eta^{ab}$ is the inverse Minkowski
metric, one can write (13) in a covariant way as
\begin{equation}
_m S_a^{~b} (P;v^l) = \bigl(2{\hat v}^l{\hat v}^m -
{\hat g}^{lm}\bigr)\nabla_{(l}\nabla_{m)} {\hat T}_a^{~b}.
\end{equation}
The sign $\dot =$ means that an equality is valid only in some special
coordinates.

The matter superenergy tensor $_m S_a^{~b}(P;v^l)$ {\it is symmetric}.

As a result of an averaging the tensor $_m S_a^{~b}(P;v^l)$, and in
consequence the averaged tensor $<_m T_a^{~b}(P;v^l)>$, do not satisfy any
local conservation laws in general relativity. However, these tensors satisfy
trivial local conservation laws\footnote{Trivial local conservation laws because the
integral superenergetic quantities or, equivalently, integral averaged
relative energy-momentum calculated from them for a closed system in special relativity vanish.} in special
relativity (see, e.g., [1-8]).

Now let us take the gravitational field and make the expansion
\begin{eqnarray}
_E t_i^{~k}(y) &=& {\alpha\over 9}\biggl[{\hat
B}^k_{~ilm} + {\hat P}^k_{~ilm}\nonumber\\
&-& {\delta_i^k\over 2}{\hat R}^{abc}_{~~~l}\bigl({\hat R}_{abcm} +
{\hat R}_{acbm}\bigr) + 2\beta^2\delta_i^k{\hat E}_{(l\vert g} {\hat
E}^g_{~\vert m)}\nonumber\\
&-& 3\beta^2{\hat E}_{i(l\vert}{\hat E}^k_{~\vert m)} + 2\beta{\hat
R}^k_{~(gi)(l\vert}{\hat E}^g_{~\vert m)}\biggr] y^ly^m + R_3.
\end{eqnarray}
Here $_E t_i^{~k}$ mean the components of the canonical Einstein
energy-momentum pseudotensor of the gravitational field.

In a holonomic frame we have
\begin{eqnarray}
_E t_i^{~k}&=&\alpha\Bigl\{\delta^k_i
g^{ms}\bigl(\Gamma^l_{mr}\Gamma^r_{sl}-\Gamma^r_{ms}\Gamma^l_{rl}\bigr)\nonumber\\
&+& g^{ms}_{~~,i}\bigl[\Gamma^k_{ms}-1/2\bigl(\Gamma^k_{tp} g^{tp}-
\Gamma^l_{tl} g^{kt}\bigr)g_{ms}\nonumber\\
& -&1/2\bigl(\delta^k_s \Gamma^l_{ml} + \delta^k_m
\Gamma^l_{sl}\bigr)\bigr]\Bigr\}.
\end{eqnarray}
\begin{equation}
\alpha = {c^4\over 16\pi G} = {1\over 2\beta},~E_i^{~k} := T_i^{~k} -
1/2\delta_i^k T,
\end{equation}
and, in any frame,
\begin{equation}
B^b_{~alm} := 2R^{bik}_{~~~(l\vert}{}R_{aik\vert m)}
-1/2\delta^b_a{}R^{ijk}_{~~~l} R_{ijkm},
\end{equation}
is the {\it Bel-Robinson tensor}, while the tensor
\begin{equation}
P^b_{~alm} := 2 R^{bik}_{~~~(l\vert}{}R_{aki\vert m)} -1/2\delta_a^b{}
R^{ijk}_{~~~l} R_{ikjm}
\end{equation}
is very closely related to the former\footnote{Very closely related because
this tensor has almost the same analytic form as the Bel-Robinson tensor and
the same symmetry properties.}.

The expansion (15) with the help of (9) and (10) gives the following
averaged gravitational relative energy-momentum tensor
\begin{equation}
<_g t_a^{~b}(P;v^l)> = _g S_a^{~b}(P;v^l){L^2\over 6},
\end{equation}
where the tensor $_g S_a^{~b}(P;v^l)$ is the {\it canonical superenergy
tensor} for the gravitational field [1-8].

We have [1-8]
\begin{eqnarray}
_g S_a^{~b}(P;v^l) &=& {2\alpha\over 9}\bigl(2{\hat v}^l{\hat v}^m -
{\hat g}^{lm}\bigr)\biggl[{\hat B}^b_{~alm} + {\hat
P}^b_{~alm}\nonumber\\
&-& 1/2\delta_a^b{\hat R}^{ijk}_{~~~m}\bigl({\hat R}_{ijkl} + {\hat
R}_{ikjl}\bigr) + 2\beta^2\delta_a^b{\hat E}_{(l\vert g}{\hat
E}^g_{~\vert m)}\nonumber\\
&-& 3\beta^2{\hat E}_{a(l\vert}{\hat E}^b_{~\vert m)} + 2\beta{\hat
R}^b_{~(ag)(l\vert} {\hat E}^g_{~\vert m)}\biggr].
\end{eqnarray}

In vacuum the tensor $_g S_a^{~b}(P;v^l)$ reduces to the simpler form
\begin{equation}
_g S_a^{~b} (P;v^l) = {8\alpha\over 9}\bigl(2{\hat v}^l{\hat v}^m -
{\hat g}^{lm}\bigr)\biggl[{\hat R}^{b(ik)}_{~~~~~(l\vert}{\hat
R}_{aik\vert m)} -1/2\delta_a^b{\hat R}^{i(kp)}_{~~~~~(l\vert}{\hat
R}_{ikp\vert m)}\biggr],
\end{equation}
which is symmetric and the quadratic form $_g S_{ab}(P;v^l){\hat
v}^a{\hat v}^b$ is {\it positive-definite}.

In vacuum we also  have the {\it local conservation laws}
\begin{equation}
\nabla_b~{_g {\hat S}_a^{~b}} = 0.
\end{equation}
and the analogous laws satisfied by  the averaged tensor $<_g
t_a^{~b}(P;v^l)>$.

The averaged energy-momentum tensors $<_m T_a^{~b}(P;v^l)>$ and $<_g
t_a^{~b}(P;v^l)>$ can be considered as the {\it averaged tensors of the relative
energy-momentum}. They can also be interpreted as the {\it fluxes} of the
appropriate canonical superenergy. It is easily seen from the
formulas (12) and (20).

Now let us consider the {\it averaged angular momentum tensors} in {\bf
GR}.  The constructive definition of these tensors, in analogy to the
definition of the averaged energy-momentum tensors, is as follows.

In normal coordinates {\bf NC(P)} we define
\begin{equation}
<M^{(a)(b)(c)}(P)> = <M^{abc}(P)> :=\displaystyle\lim_{\varepsilon\to
0}{ \int\limits_{\Omega}{\bigl[M^{(a)(b)(c)}(y) -
M^{(a)(b)(c)}(P)\bigr]d\Omega}\over\varepsilon^2/2\int\limits_{\Omega}d\Omega},
\end{equation}
where
\begin{equation}
M^{(a)(b)(c)}(y) :=
M^{ikl}(y){}e_i^{~(a)}(y){}e_k^{~(b)}(y){}e_l^{~(c)}(y),
\end{equation}
\begin{equation}
M^{(a)(b)(c)}(P) := M^{ikl}(P)
e_i^{~(a)}(P){}e_k^{~(b)}(P){}e_l^{~(c)}(P) =
M^{ikl}(P)\delta_i^a\delta_k^b\delta_l^c = M^{abc}(P),
\end{equation}
are the {\it physical} (or tetrad) components of the field $M^{ikl}(y)=
(-)M^{kil}(y)$ which describes the angular momentum densities
\footnote{ Of course, $M^{abc}(P) = 0$, but we leave $M^{abc}(P)$ in our
formulas.}.
As in (2) and (3) , $e^i_{~(a)}(y),~~e_k^{~(b)}(y)$ denote mutually dual
orthonormal tetrads parallelly propagated along geodesics through {\bf
P} such that $e^i_{~(a)}(P) = \delta^i_a,~~e_k^{~(b)}(P) = \delta_k^b$.
The compact four-dimensional domain $\Omega$ is defined in the same way
as in the formula (1) and we will again take $\Omega$ as a sufficiently small
four-dimensional ball with centre at {\bf P} and with radius $R = \varepsilon
L$.

At {\bf P} the tetrad and normal components of an object are equal. We
apply this once more and omit tetrad brackets for the indices of any
quantity attached to the point {\bf P}; for example, we write
$M^{abc}(P)$ instead of $M^{(a)(b)(c)}(P)$ and so on.

For matter as $M^{ikl}(y)$ we take
\begin{equation}
_m M^{ikl}(y) =\sqrt{\vert g\vert}\bigl[y^i T^{kl}(y) - y^k
T^{il}(y)\bigr],
\end{equation}
where $T^{ik}(y) = T^{ki}(y)$ are the components of a symmetric
energy-momentum tensor of matter and $y^i$ denote the normal coordinates
{\bf NC(P)}.

The formula (27) gives us the total angular momentum densities, orbital and
spinorial, because the symmetric energy-momentum tensor of matter
$T^{ik} = T^{ki}$ comes from the canonical one by
using the {\it Belinfante-Rosenfeld} symmetrization procedure and,
therefore, includes the canonical spin of matter [14].

For the gravitational field we take the gravitational angular momentum
pseudotensor proposed by Bergmann and Thomson [14,18] which  in a
{\bf NC(P)} (and in any other holonomic frame) reads
\begin{equation}
_g M^{ikl}(y) = _F U^{i[kl]}(y) - _F U^{k[il]}(y)+\sqrt{\vert
g\vert}\bigl(y^i _{BT} t^{kl} - y^k _{BT} t^{il}\bigr),
\end{equation}
where, in a holonomic frame,
\begin{equation}
_F U^{i[kl]}:= g^{im}{}_F U_m^{~[kl]} =\alpha
g^{im}{g_{ma}\over\sqrt{\vert g\vert}}\biggl[(-g)\bigl(g^{ka} g^{lb} -
g^{la} g^{kb}\bigr)\biggr]_{,b}
\end{equation}
are {\it Freud's superpotentials} with the first index raised and
\begin{equation}
_{BT} t^{kl}:= g^{ki}{} _Et_i^{~l} + {g^{mk}_{~~,p}\over\sqrt{\vert
g\vert}} {} _F U_m^{~[lp]}
\end{equation}
are the components of the {\it Bergmann-Thomson} gravitational
energy-momentum pseudotensor [14,18]. $_E t_i^{~k}$ mean the components of the
{\it Einstein canonical gravitational energy-momentum pseudotensor} of the gravitational
field.

The Bergmann-Thomson gravitational angular pseudotensor is most closely
related to the Einstein canonical energy-momentum complex $_E K_i^{~k}:=
\sqrt{\vert g\vert}\bigl(T_i^{~k} + _E t_i^{~k}\bigr)$, matter and
gravitation,  and it has better physical and transformational properties than the
famous gravitational angular momentum pseudotensor proposed by Landau and
Lifschitz [15-17]. This is why we apply it here.

One can interpret the Bergmann-Thomson gravitational angular momentum
pseudotensor as the sum of the {\it spinorial part}
\begin{equation}
S^{ikl} := _F U^{i[kl]} - _F U^{k[il]}
\end{equation}
and the {\it orbital part}
\begin{equation}
O^{ikl} := \sqrt{\vert g\vert}\bigl(y^i{}_{BT} t^{kl} - y^k{}_{BT}
t^{il}\bigr)
\end{equation}
of the gravitational angular momentum ``densities''.

Substitution of (27) and (28) (expanded up to third order), (9),(10) and the
expansion
\begin{equation}
\sqrt{\vert g\vert} = 1 -1/6{\hat R}_{ab}y^ay^b + R_3 = 1-1/6\beta{\hat
E}_{ab}y^ay^b + R_3,
\end{equation}
into (24) gives us the following {\it averaged angular momentum
tensors} for matter and gravitation respectively
\begin{equation}
<_m M^{abc}(P;v^l)> = _m S^{abc}(P;v^l){L^2\over 6},
\end{equation}
\begin{equation}
<_g M^{abc}(P;v^l)> = _g S^{abc}(P;v^l){L^2\over 6}.
\end{equation}

Here
\begin{equation}
_m S^{abc}(P;v^l) = 2\bigl[\bigl(2{\hat v}^a{\hat v}^p - {\hat
g}^{ap}\bigr)\nabla_p{\hat T}^{bc} - \bigl(2{\hat v}^b{\hat v}^p - {\hat
g}^{bp}\bigr) \nabla_p {\hat T}^{ac}\bigr],
\end{equation}
and
\begin{eqnarray}
_g S^{abc}(P;v^l) &=& \alpha\bigl(2{\hat v}^p{\hat v}^t - {\hat
g}^{pt}\bigr)\biggl[\beta\bigl({\hat g}^{ac}{\hat g}^{br} -{\hat
g}^{bc}{\hat g}^{ar}\bigr)\nabla_{(t}{\hat E}_{pr)}\nonumber\\
&+& 2{\hat g}^{ar}\nabla_{(t}{\hat R}^{(b}_{~~p}{}^{c)}_{~~r)} - 2{\hat
g}^{br} \nabla_{(t}{\hat R}^{(a}_{~~p}{}^{c)}_{~~r)}\nonumber\\
&+& 2/3{\hat g}^{bc}\bigl(\nabla_r{\hat R}^r_{~(t}{}^{a}_{~p)}
-\beta\nabla_{(p} {\hat E}^a_{~t)}\bigr) -2/3{\hat g}^{ac}\bigl(\nabla_r
{\hat R}^r_{~(t}{}^b_{~p)} - \beta\nabla_{(p} {\hat
E}^b_{~t)}\bigr)\biggr]
\end{eqnarray}
are the components of the {\it canonical angular supermomentum tensors} for
matter and gravitation, respectively [4,6,8].

In special relativity the averaged tensor $<_m
M^{abc}(P;v^l)>$, and the canonical angular supermomentum tensors for
matter $_m S^{abc}(P;v^l)$ satisfy trivial conservation laws [1-8].
In the framework of the {\bf GR} only the tensors $_g S^{abc}(P;v^l)$ and $<_g M^{abc}(P;v^l)>$
satisfy local conservation laws in vacuum.

In vacuum, when $T_{ik} = 0 \Longleftrightarrow E_{ik} := T_{ik} -1/2
g_{ik}T = 0$, the canonical gravitational angular supermomentum tensor
$_g S^{abc}(P;v^l) = (-) _g S^{bac}(P;v^l)$ given by (37) simplifies to
\begin{equation}
_g S^{abc}(P;v^l) =2\alpha\bigl(2{\hat v}^p{\hat v}^t - {\hat
g}^{pt}\bigr)\biggl[{\hat g}^{ar}\nabla_{(p}{\hat R}^{(b}_{~~t}{}^{c)}_{~~r)}
-{\hat g}^{br}\nabla_{(p} {\hat R}^{(a}_{~~t}{}^{c)}_{~~r)}\biggr].
\end{equation}

Some remarks are in order:
\begin{enumerate}
\item The orbital part $O^{ikl} =\sqrt{\vert g\vert}\bigl(y^i_{BT}
t^{kl} - y^k _{BT} t^{il}\bigr)$ of the $_g M^{ikl}$ {\it does not
contribute} to the tensor $_g S^{abc}(P;v^l)$ and, therefore, also to the
tensor $<_g M^{abc}(P;v^l)>$. Only the spinorial part $S^{ikl} = _F U^{i[kl]} - _F
U^{k[il]}$ gives nonzero contribution to these tensors.
\item The averaged angular nomentum tensors $<_gM^{abc}(P;v^l)>, ~~<_m
M^{abc}(P;v^l)>$, like as the canonical angular supermomentum tensors,
{\it do not need} any radius-vector for existing.
\end{enumerate}

The averaged tensors $< _m M^{abc}(P;v^l)>,~~<_g M^{abc}(P;v^l)>$,
likely as the averaged relative energy-momentum tensors, can be interpreted as
the {\it averaged tensors of the relative angular momentum}\footnote{Of
course, the angular momentum is always relative quantity, in
principle. Despite that we will keep the term {\it relative angular
momentum tensors}.} and also as the {\it fluxes} of the appropriate angular
supermomentum.

The formulas (12),(20),(34) and (35) give the direct link beteween
the canonical superenergy and angular supermomentum tensors
\begin{equation}
 _gS_a^{~b}(P;v^l),~_m S_a^{~b}(P;v^l), ~_g S^{abc}(P;v^l), ~~_m
S^{abc}(P;v^l)
\end{equation}
and the averaged relative energy-momentum and angular momentum tensors
\begin{equation}
<_g t_a^{~b}(P;v^l)>, <_m T_a^{~b}(P;v^l)>, <_g
M^{abc}(P;v^l)>, <_m S^{abc}(P;v^l)>.
\end{equation}
Namely, it is easily seen from these formulas that the averaged relative energy-momentum
and angular momentum tensors {\it differ} from the canonical superenergy and angular supermomentum
tensors {\it only} by the constant scalar multiplicator ${L^2\over 6}$, where $L>0$
means some fundamental length. Thus, from the mathematical point of view, these two
kind of tensors are equivalent. Physically they {\it are not} because their
components have different dimension. Moreover the averaged
energy-momentum and angular momentum tensors depend on a fundamental
length $L>0$, i.e., they need introduction a supplementary element into
{\bf GR}\footnote{The fundamental length $L>0$ must be infinitesimally
small because its existence violates local Lorentz invariance. It is generally
belived that a fundamental length exists in Nature.}. Owing to the last fact and the formulas (12),(20), (34), (35) it seems
that the canonical superenergy and angular supermomentum tensors are {\it
more fundamental} than the averaged energy-momentum and angular momentum
tensors.
But the averaged energy-momentum and angular momentum tensors have one
important superiority over the canonical superenergy and angular
supermomentum tensors: their components {\it possesse proper dimensions} of
the energy-momentum and angular momentum densities.

The averaged tensors
\begin{equation}
<_g t_a^{~b}(P;v^l)>, ~<_m T_a^{~b}(P;v^l)>,~~ <_g
M^{abc}(P;v^l)>, ~<_m M^{abc}(P;v^l)>
\end{equation}
depend on the four-velocity ${\vec v}$ of a fiducial observer {\bf O} which is at rest at
the beginning {\bf P} of the normal coordinates {\bf NC(P)} used for
averaging and on some fundamental length $L>0$. After fixing the
fundamental length $L$ one can determine univocally these tensors along the world line
of an observer {\bf O}.

In general one can {\it unambiguously determine} these tensors (after fixing
$L$) in the whole spacetime or in some domain $\Omega$ if in the spacetime or
in the domain $\Omega$ a geometrically distinguised timelike unit vector
field ${\vec v}$ exists. An example of such a kind of the spacetime is
given by Friedman universes.

One can try to fix\footnote{But this is {\it not necessary}. One can effectively use
the averaged energy-momentum and angular momentum tensors {\it without fixing
L} explicitly.} the fundamental length $L$, e.g., by using loop quantum gravity.
Namely, one can take as $L$ the smallest length $l$ over which the
classical model of the spacetime is admissible.
\par
Following loop quantum gravity [19-29] one can say about continuous classical
differential geometry already just a few orders of magnitude above the Planck
scale, e.g., for distances $l\geq 100L_P = 100\sqrt{{G\hbar\over c^3}}
\approx 10^{-33}$ m. So, one can take as the fundamental length $L$ the
value $L = 100 L_P \approx 10^{-33}$ m.\footnote{Concerning other
propositions fixing of $L$ see, e.g., [9--12].}

After fixing the fundamental length $L$ one has the averaged relative
energy-momentum and angular momentum tensors as precisely defined as
the canonical superenergy and angular supermomentum tensors are.

The averaged tensors (with $L$ fixed or no)
\begin{equation}
<_mT_a^{~b}(P;v^l)>, <_g t_a^{~b}(P;v^l)>, <_m M^{abc}(P;v^l)>, <_g
M^{abc}(P;v^l)>
\end{equation}
give us as good tool to a local analysis ( and also to global analysis iff in spacetime
a privileged global unit timelike vector field exists))
of the gravitational and matter fields as the canonical superenergy and
angular supermomentum tensors
\begin{equation}
_m S_a^{~b}(P;v^l), ~_g S_a^{~b}(P;v^l),
~_m M^{abc}(P;v^l), ~_g M^{abc}(P;v^l)
\end{equation}
give. For example, one can apply the averaged energy-momentum and
angular momentum tensors to the all problems which have been analyzed in the
papers [1-8] by using the canonical superenergy and angular
supermomentum tensors.
\section{Some applications of the averaged relative energy-momentum tensors}
In this paper we apply the averaged gravitational relative energy-momentum tensor\hfill\break
$<_g t_a^{~b}(P;v^l)>$ only to decide if free vacuum
gravitational field has energy-momentum; especially, if
gravitational waves carry any energy-momentum, and the averaged
gravitational and matter relative energy-momentum tensors to analyze the energy
and momentum of the Friedman universes.

Albrow and Tryon were the first who assumed that the net energy
of the closed Friedman universes may be equal to zero [30-31]. We
will show in this paper that this assumption is, most probably, {\it incorrect}.

Let us begin from the vacuum gravitational energy and momentum. The problem
was revived recently because some authors conjectured [32-36], by using
coordinate dependent\footnote{By ``coordinate dependent'' quantity we
mean a quantity which is not a tensor (in general--which is not a tensor
valued p-form). By ``coordinate independent'' quantity we mean a tensor
quantity (in general -- a tensor valued p-form).} pseudotensors and
double index complexes, that the energy and momentum in general relativity are confined only to the regions of
non-vanishing energy-momentum tensor of matter and that the gravitational waves carry
no energy and momentum.
The argumentation is the following. For some solutions to the Einstein
equations and in some special coordinates, e.g., in Bonnor's spacetime
[37] in Bonnor's or in Kerr-Schild coordinates, the Einstein canonical
gravitational energy-momentum pseudotensor (and other most frequently
used gravitational energy-momentum pseudotensors also) {\it globally vanishes} outside of the domain in which
$T^{ik}\not= 0$. The analogous global vanishing of the canonical
pseudotensor $_E t_a^{~b}$ we have for the plane and for the
plane-fronted gravitational waves in, e.g., null coframe [3,38].
But one should emphasize that all these results are {\it coordinate
dependent} [3,7,38], i.e., in {\it other coordinates} the used gravitational
energy-momentum pseudotensors {\it do not vanish} in vacuum. Moreover, one
should interpret physically the global vanishing of the canonical pseudotensor (and other pseudotensors
also) in some coordinates in vacuum as a {\it global cancellation} of the energy-momentum of the real
gravitational field which has $R_{iklm}\not= 0$ with energy-momentum of the inertial forces field
which has $R_{iklm}=0$; {\it not as a proof of vanishing of the energy-momentum  of the real
gravitational field}. It is because the all used pseudotensors were entirely constructed from
the Levi-Civita's connection $\Gamma^i_{~kl} = \Gamma^i_{~lk}$ and from
the metric $g_{ik}$ which describe a mixture of the
real gravitational field ($R_{iklm}\not= 0$) and an inertial forces field ($R_{iklm}= 0$).

In order to get the coordinate independent results about
energy-momentum of the {\it the real gravitational field}  one must use
tensorial expressions which depend on curvature tensor, like the averaged
gravitational relative energy-momentum tensor $<_g t_a^{~b}(P;v^l)>$. This tensor
vanishes iff $R_{iklm}=0$, i.e., iff the spacetime is flat and we have no real
gravitational field.

When calculated, the averaged gravitational relative energy-momentum
tensor\hfill\break  $<_g t_a^{~b}(P;v^l)>$ always gives the {\it positive-definite} averaged free
relative gravitational energy density and, in the case of a gravitational wave, its non-zero
flux. It is easily seen from the our papers [1-8,38] in which we have
used the canonical gravitational superenergy tensor and from the
formula (20) of this paper which gives the direct connection
between the averaged relative gravitational energy-momentum tensor
and the canonical gravitational superenergy tensor.

Thus, the conjecture about localization of the gravitational energy only
to the regions of the non-vanishing energy-momentum tensor of matter
{\it is incorrect} for the real gravitational field which has
$R_{iklm}\not= 0$.

It is interesting that the gravitational angular momentum pseudotensor
(28) {\it does not vanish} in Bonnor's spacetime and in Bonnor's coordinates
{\it outside} of the domain in which $T^{ik}\not= 0$. This important
fact which, as I think, is unknown for the authors of the conjecture,
gives other {\it direct proof} that this conjecture {\it is incorrect}. If
the conjecture were correct, then we would have an absurd situation: the
energy-momentum density--free vacuum gravitational field has non-vanishing
``densities'' of the angular momentum.

In a similar way as above one can use the averaged gravitational relative
angular momentum tensor $<_g M^{abc}(P;v^l)>$ to coordinate independent
analysis of the angular momentum of the real gravitational field.

Now, let us pass to the problem of the energy and momentum of the
Friedman universes. Of course, the problem of the global energy and
global linear (or angular) momentum for Friedman universes (and also
for more general universes) is not {\it well-posed} from the physical
point of view because these universes are not
asymptotically flat spacetimes [39]. Despite this important fact
recently many authors concluded [40-50] that the energy and momentum of the
Friedman universes, flat and closed, are equal to zero locally and globally (flat universes) or
only globally (closed universes). Such conclusion, which has a mathematical sense,
originated from calculations performed in special comoving coordinates called ``Cartesian
coordinates'' by using {\it coordinate dependent} double index
energy-momentum complexes, matter and gravitation.

One can introduce in {\bf GR} many different energy-momentum complexes. The six of them
are most frequently used: Einstein's canonical complex, Landau-Lifshitz complex,
Bergmann-Thomson complex, M\o ller complex, Papapetrou complex and
Weinberg energy-momentum complex. These all energy-momentum complexes
{\it are neither geometrical objects nor coordinate independent objects}, e.g., they can vanish
in some coordinates locally or globally and in other coordinates they can be
different from zero. It results that the double index energy--momentum
complexes and the gravitational energy-momentum pseodotensors {\it have
no physical meaning} to a local analysis of the gravitational
field, e.g., to study gravitational energy-density distribution.
They can be reasonably used {\it only to calculate the global quantities}
for the very precisely defined asymptotically flat spacetimes (in spatial
or in null direction).

The general opinion is that the best one of the all possible double
index energy-momentum complexes from physical and geometrical points of view is the
canonical Einstein's double index energy-momentum complex $_E K_i^{~k} =
\sqrt{\vert g\vert}\bigl(T_i^{~k} + _E t_i^{~k}\bigr)$. The global results obtained
by use of this canonical energy-momentum complex are usually
treated as correct and giving some pattern. In fact, the other double index
energy-momentum complexes were constructed following the instruction: they should
give the same global results as the Einstein energy-momentum complex gives at least
in the simplest cases, e.g., in the case of a closed system. That is why we
have confined in the paper (and also in the all our previous papers) only to
this double index energy-momentum complex.

So, let us consider the results of the formal calculations of the global
energy and momentum for Friedman universes in the standard comoving
coordinates by using canonical Einstein's double index energy-momentum
complex. Any other sensible double index energy-momentum complex gives equivalent
results.
\begin{enumerate}
\item In the ``Cartesian coordinates'' $(t,x,y,z)$ in which the line
element has the form\footnote{From now on we will use {\it geometrized
units} in which $G = c =1$.}
\begin{equation}
ds^2 = dt^2 - R^2(t){(dx^2+dy^2+dz^2)\over[1+k/4(x^2+y^2+z^2)]^2},~~~k =
0,^+_-1,
\end{equation}
we obtain after simple calculations [1,5] that for flat universes the global
quantities $P_i ~(i = 0,1,2,3)$, where $P_i$ mean the components of the
energy-momentum contained inside of a slice $t = const$, are equal to zero. In
this case the all integrands (energy and momentum ``densities'') in the integrals on $P_i~~(i = 0,1,2,3)$
identically vanish because they are multiplied by the curvature index $k$. So, one can say
that for flat Friedman universes the integral quantities $P_i ~~(i=0,1,2,3)$
vanish locally and globally in the ``Cartesian'' coordinates\footnote{It is
interesting that the angular momentum ``densities'' when calculated, e.g., by
using Bergmann--Thomson angular momentum complex (28) do not vanish in
the case even for flat Friedman universes.}.

For closed Friedman universes we also get $P_i = 0,~~(i=0,1,2,3)$, but this time
the integrands do not vanish. Only after integration one gets that the integrals
representing $P_i, (i =0,1,2,3)$ are equal to zero. In the case of the open Friedman universes
one gets $E = P_0 = (-)\infty, ~P_1 = P_2 = P_3 = 0$. The integrands
also do not vanish in this case.

\item In the coordinates $(t,\chi,\vartheta,\varphi)$ in which the line
element reads
\begin{equation}
ds^2 = dt^2 -R^2(t)[d\chi^2 + S^2(\chi)(d\vartheta^2 + \sin^2\vartheta
d\varphi^2)],
\end{equation}
where
\begin{equation}
S(\chi) =\bigl\{sin\chi~~if~~ k = 1, ~~\chi~~ if~~ k=0,~~
sh\chi~~ if~~ k =-1\bigr\},
\end{equation}
one gets drastically different results: $E = P_0 =(-)\infty, ~~ P_1 =
(-)\infty,~~P_2 = P_3 = 0$ for flat universes; $ E = P_0 = {\pi\over 2}
R(t), ~~P_1 = P_2 = P_3 =0$ for closed univeres and  $E = P_0
=(-)\infty, ~~ P_1 =(-)\infty, ~~P_2 = P_3 =0$ for open universes.
\item  Finally, in the coordinates $(t,r,\vartheta,\varphi)$ in which the
line element has the form
\begin{equation}
ds^2 = dt^2 - R^2(t)[{dr^2\over(1-kr^2)} +
r^2(d\vartheta^2+\sin^2\vartheta d\varphi^2)],~~k=0,^+_-1,
\end{equation}
we obtain the following results: $E = P_0 =(-)\infty,~~ P_1 = (-)\infty,
~~P_2 = P_3 = 0$ for flat universes; $E = P_0 = {\pi\over 4}R(t),~~P_1 =
(-)\infty,~~P_2 = P_3 = 0$ for closed universes and $E= P_0 =(-)\infty,~~ P_1
=(-)\infty,~~P_2 = P_3 =0$ for open Friedman universes.
\end{enumerate}

In the all cases in which the integrands (=``densities'' of the calculated
four-momentum) do not vanish, these integrands go to zero if
$R(t)\longrightarrow 0$. So, these integrands (``densities''of the energy-momentum) are not suitable for
analysis of the Big-Bang singularity.

The authors which assert that the energy and momentum of the Friedman
universes, flat and closed, are equal to zero have performed their
calculations only in the ``Cartesian'' comoving coordinates $(t,x,y,z)$
by using coordinate dependent double index energy-momentum complexes and have got
zero results. But in the case of the Friedman universes the
``Cartesian''coordinates {\it are by no means better} than the comoving coordinates $(t,\chi,\vartheta,\varphi)$ or
$(t,r,\vartheta,\varphi)$ in which we have obtained non-zero results. Only in a flat and
in an asymptotically flat spacetimes one can distinguish in some reasonable way
the Cartesian coordinates; but {\it not in the case of the Friedman universes}. So,
the conclusion of these authors about vanishing of the
energy and linear momentum of the Friedman universes, flat and closed,
{\it cannot be correct}.

By using double index energy-momentum complexes one rather should conclude that the
energy and momentum of the Friedman universes explicite depend on the
used comoving coordinates and, therefore, that {\it they are undetermined}
locally and globally. This last conclusion is very sensible because {\it
one cannot measure the global energy and global momentum of the Friedman
(and more general) universes}. One can do this only in the case of an isolated system [39]. On
the other hand the former conclusion directly follows from the coordinate dependence of
the energy-momentum complexes.

May be one would try to support the {\it mathematically sensible} hypothesis which states that energy and
momentum of the Friedman universes, flat and closed, disappear by using
coordinate independent expressions, like Pirani's expression on global
energy, matter and gravitation, or like single index Komar's expression
(Komar's single index complex) on global energy-momentum and global angular momentum,
matter and gravitation \footnote{We would like to remark that the
Pirani's and Komar's expressions, though coordinate independent, depend
(like double index energy-momentum complexes) not only on real
gravitational field ($R_{iklm}\not= 0$) but also on inertial forces field ($R_{iklm}
= 0$).}.

The Pirani's expression (for the energy only, see, eg., [51]) is unique
and can be applied in a spacetime having a privileged set of observers whose
world-lines form a normal congruence. In such spacetime there exists a family of
spatial hypersurfaces which are orthogonal to the four-velocities of this set
of observers.

The Pirani's expression is coordinate independent but it has two
defects: calculated total energy density, matter and gravitation, {\it is not
positive-definite}, and, if the congruence is geodesic, then the total
energy-density {\it is identically zero}, and, in consequence, the global
energy {\it trivially vanishes in the case}. However, this zero values
{\it are not a property of the gravitational and matter fields}. They
are only a property of the  geodesics congruence.

In Friedman universes does exist privileged set of observers called {\it
fundamental or isotropic} observers. For these observers the
four-velocity ${\vec v}$ has components $v^k = \delta^k_0$ in a comoving
coordinates and the family of the spatial hypersurfaces orthogonal to ${\vec
v}$ is given by $t = const.$ But, unfortunately, the congruence of the
isotropic observers in Friedman universes is geodesic and, therefore,
the Pirani's expression {\it fails} in the case {\it giving trivially
zero}.

On the other hand, coordinate independent Komar's expression (see, e.g.,
[51-53]) {\it needs Killing vector fields}: translational timelike
Killing vector field as {\it energy descriptor}, translational spatial Killing
vector fields as {\it descriptors of the linear momentum} and rotational
spatial Killing vector fields as {\it descriptors of the angular
momentum}.

Friedman universes admit only six linearly independent spatial Killing vector
fields, three translational Killing vector fields and three rotational
Killing vector fields (see, e.g., [54]). So, one can consider in Friedman universes
six coordinate independent integrals (scalars) which correctly represent (from mathematical point of view)
the components of the global linear momentum and the components of the global
angular momentum (see, eg., [54]). These integrals {\it trivially} vanish for
Friedman universes, i.e., integrands in these integrals {\it identically}
vanish, independently of the curvature index $k=0,^+_-1$. This is very
sensible result and it can be interpreted as a mathematically correct proof that the
linear and angular momentum for Friedman universes disappear in a
comoving coordinates.

But we still have a problem with energy of the Fiedman universes {\it because
we have no energy descriptor}, i.e., translational timelike Killing
vector field, in these universes. Therefore, one cannot use the
coordinate independent Komar's expression in order to calculate
correctly from the mathematical point of view
the energy of the Friedman universes.

If one formally uses in Komar's expression the four-velocity of the privileged set of the isotropic
observers as the energy descriptor, then {\it one will get identically zero} because for a
geodesic timelike congruence the integrand in this expression, like
integrand in Pirani's expression, identically vanishes. But this vanishing is
also only a property of the geodesics congruence. It is not a property of the
gravitational and matter fields.

Resuming, one cannot use the coordinate independent Pirani's and Komar's
expressions in order to correctly prove\footnote{Correctly from the mathematical point of view.}
the statement that the energy of the Friedman universes disappears, i.e., that these universes are complete
energetic nonentity.

For this purpose one cannot also use the coordinate independent KBL bimetric
approach [55] because the results obtained in this approach depend not only on
the used background but also on mapping of the real spacetime onto this background.

Therefore, the {\it mathematically sensible}  statement that the closed and flat Friedman universes have
no energetic content {\it is still not satisfactory proved}.

It is interesting that the using of the coordinate independent averaged
relative energy-momentum tensors to analyze the energetic content of the
Friedman universes lead us to {\it positive-definite results} for the all
Friedman universes.

Namely, let us apply the averaged relative energy-momentum tensors for gravitation $<_g t_i^{~k}(P;v^l)>$
and for matter $<_m T_i^{~k}(P;v^l)>$ to calculate the averaged relative energy density for Friedman
(and more general) universes. With this aim let us define
\begin{equation}
_g\epsilon := <_g t_a^{~b}(P;v^l)>v^av_b
\end{equation}
----- the averaged relative gravitational energy density,
\begin{equation}
_m\epsilon:= <_m T_a^{~b}(P;v^l)>v^av_b
\end{equation}
----- the averaged relative matter energy density,
and
\begin{equation}
\epsilon:= _g\epsilon + _m \epsilon
\end{equation}
----- the averaged relative total energy density.

Here $v^a$ are the components of the four-velocity of an observer {\bf
O} which is studying gravitational and matter fields.

In Friedman universes, if we take as the observers {\bf O} the globally
defined set of the fundamental observers, then we can also define the
global averaged total relative energy $E$ of a  Friedman universe
\begin{equation}
E := \int\limits_{t = const}\epsilon\sqrt{\vert g\vert} d^3v ~{\dot
=}\int\limits_{t = const} \bigl[<_g t_i^{~0}> + <_m T_i^{~0}>\bigr]
v^i\sqrt{\vert g\vert} d^3v,
\end{equation}
and, in analogous way, the global averaged relative energy for matter
and for gravitation.

Here $d^3v$ means the product of the diferentials of the
coordinates which parametrize   slices $t = const$ of the Friedman
universes, e.g., $d^3v = dxdydz$ in the Cartesian comoving
coordinates $(t,x,y,z)$.

After something tedious but very simple calculations we will obtain for Friedman universes
[1,2,5]\footnote{The results given below are easily seen from the our previous papers [1,2,5]
and from the formulas (12) and (20) which connect the canonical superenergy tensors used in the
papers [1,2,5] with the averaged relative energy and momentum tensors which we
are using in this paper.}:
\begin{enumerate}
\item $_g\epsilon$, $_m\epsilon$ and, in consequence $\epsilon$, are
{\it positive definite} for the all Friedman universes.
\item $\displaystyle\lim_{R\to 0}{}_g\epsilon = \displaystyle\lim_{R\to 0}
{}_m\epsilon =\displaystyle\lim_{R\to 0}{}\epsilon = +\infty, ~~(k =
0,^+_- 1)$.

It follows from this that one can use the averaged relative energy densities to study the
Big-Bang singularity.
\item $\displaystyle\lim_{R\to\infty}{} _g\epsilon =
\displaystyle\lim_{R\to\infty}{} _m \epsilon =
\displaystyle\lim_{R\to\infty}{}\epsilon  = 0, ~~(k = 0,-1)$.
\item The global averaged relative energies, gravitation, matter and total, are infinite
($+\infty$) for flat and for open Friedman universes and they are finite and positive
for closed Friedman universes.
\end{enumerate}

Also the other three invariant integrals which formally represent the components
$P_{(\alpha)} ~~(\alpha = 1,2,3)$ of the global averaged relative linear momentum for
Friedman universes
\begin{equation}
P_{(\alpha)}:= \int\limits_{t = const}\bigl\{<_g t_i ^{~0}> + < _m
T_i ^{~0}>\bigr\}e^i_{~(\alpha)}\sqrt{\vert g\vert}d^3v,~~(\alpha = 1,2,3),
\end{equation}
{\it vanish trivially} in a comoving coordinates [1,2,5] because the
integrands in these integrals (densities of the averaged relative linear
momentum components) {\it identically vanish} [1,2,5].

Here $e^i_{~(\alpha)}, ~~(\alpha = 1,2,3)$ mean the components of the
three translational spatial Killing vector fields (descriptors of the
linear momentum) which exist in the Friedman universes (see, e.g.,
[54]).

We would like to emphasize that the integrals (51) and (52) do not
depend on the used coordinates. They depend only on a slice $t = const$.

The all above results are very sensible and satisfactory from the
physical point of view.

We will finish this Section with remark that the analogous situation as
for flat Friedman universes one has also for the more
general, only homogeneous, Kasner vacuum universes [15] and
Bianchi--type I universes filled with stiff matter (see, e.g.,
[44-50, 56-59]). Namely, the most frequently used double-index
energy-momentum complexes, when used in Cartesian comoving
coordinates to analyze of these universes, give zero results
locally and globally.

Of course, in other comoving coordinates,
e.g., in the $t,r,\vartheta,\varphi$  comoving coordinates, we
have non-zero and globally divergent results.

If one applies the averaged relative energy-momentum tensors
$<_g t_a^{~b}(P;v^l)>, \\<_m T_a^{~b}(P;v^l)>$ to analyze of a
vacuum Kasner universe and a Bianchi--type I universe filled with
stiff matter, then one gets the following, coordinate independent
results:
\begin{enumerate}
\item The averaged relative gravitational energy of a vacuum Kasner universe has
{\it positive-definite} density and the same limits when $t\longrightarrow 0$
or when $t\longrightarrow +\infty$ as it was in the case of a flat
Friedman universe. Also the suitable integral global quantity defined in analogous
way as in the case of the Friedman universes is divergent to $+\infty$.
\item For an expanding Bianchi--type I universe filled with stiff matter
the averaged relative gravitational energy density and the averaged relative
energy-density for matter are still {\it positive-definite} and
lead to divergent to $+\infty$ global energies.
\end{enumerate}

Thus, one can conclude that these two more general,
only homogeneous universes, like Friedman flat universes,
also {\it are not energetic nonentity}.

Concerning of the components of the linear momentum for Kasner
vacuum universes and for Bianchi--type I universes filled with
stiff matter one can easily check that these components, defined
in analogous way as in the case of the Friedman universes,
{\it identically vanish locally and globally} in a comoving coordinates.

\section{Conclusion}
We have introduced in the paper the averaged tensors of the relative
energy-momentum and the averaged tensors of the relative angular
momentum, for matter and for gravitation. These tensors are very closely
related to the canonical superenergy and angular supermomentum tensors and they
can be used to analyze the same problems which we have analyzed in the our
papers [1-8] with the help of the canonical superenergy and
angular supermomentum tensors. The superiority of the averaged relative
energy-momentum and angular momentum tensors in comparison with the canonical
superenergy and angular supermomentum tensors is the following: the
averaged tensors have proper dimensionality of the energy-momentum and
angular momentum densities.

The averaged relative energy-momentum and relative angular momentum
tensors of the gravitational field {\it refer to the energy-momentum and
angular momentum of the real gravitational field} for which we have
$R_{iklm}\not= 0$. These tensors vanish iff $R_{iklm} =0$, i.e., iff
{\it we have no real gravitational field}.

In our opinion the all existing (and projected in near future) detectors
of the gravitational waves will measure the averaged relative
gravitational energy density and its flux; not the gravitational energy
defined by pseudotensors. It is easily seen from the fact that the acting of
these detectors relies on the equations of the geodesics deviation which
explicitly depend on the curvature tensor.

In this paper we have applied the averaged relative gravitational
energy-momentum tensor to decide if free vacuum gravitational field has
energy and momentum and the averaged gravitational
and matter relative energy-momentum tensors to analyze energy and momentum of
the Friedman universes and also to analyze the Kasner and Bianchi--type I universes.
The latter problem is recently very popular
despite the fact that the problem of the global quantities for Friedman
universes (and for more general cosmological models also) {\it is not well-posed from the
physical point of view}. The global energy and momentum {\it have physical
meaning} only when spacetime is asymptotically flat either in spatial or null
direction.
Of course, this is not a case of the Friedman and Kasner or Bianchi--type I cosmological models.

We have obtained the following results:
\begin{enumerate}
\item The real vacuum gravitational field for which we have $R_{iklm}\not=
0$ {\it always} possesses his own positive-definite averaged relative energy
density and in the cases in which the gravitating system is not at rest,
the gravitational field possesses also the non-zero averaged relative linear
momentum.
\item The coordinate independent averaged relative energy-momentum
tensors, gravitation and matter, give positive-definite densities of the
averaged relative energy, matter and gravitation, for the all Friedman
universes. Therefore, these tensors indicate that the Friedman universes
{\it are not energetic nonentity}. They {\it are not energetic nonentity} in the following sense:
one can construct from the canonical energy-momentum complex, matter and gravitation, non-local
tensorial, i.e., coordinate-independent expressions with correct dimensions which
give positive-definite energy densities for the all Friedman universes.

The averaged relative energy-momentum tensors tensors give also zero values
of the averaged relative linear momenta for these universes in a comoving coordinates.
\end{enumerate}

The above results directly follow from the results obtained in the our
previous papers [1-5] in which we have used the canonical superenergy (and
angular supermomentum) tensors, gravitation and matter, and from the
formulas (12) and (20) of this paper which connect the averaged relative
energy-momentum tensors with the canonical superenergy tensors.

The coordinate independent results presented in this paper for the
Friedman universes are very satisfactory from the physical
point of view. Much more satisfactory than the strange, coordinate dependent
results which one obtains by using gravitational energy-momentum
pseudotensors and double index energy-momentum complexes, matter and gravitation.
By using of these objects one can only conclude that the energy and
momentum of the Friedman universes {\it are undetermined} locally and
globally.

The analogous conclusion as given above for Friedman universes is
also correct for the more general  Kasner and Bianchi--type I
universes.

We are planning to use in a future the averaged relative
energy-momentum tensors, and also the averaged tensors of the relative
angular momentum, to analyze much more general homogeneous universes, like the universes
which have been considered in the papers [44-50, 56-60].

\end{document}